# SALSA: a new versatile readout chip for MPGD detectors

D. Neyret,[a,1] P. Baron,[a] F. Bouyjou,[a] M. Bregant,[b] S. Chevobbe,[a] Y. Degerli,[a] C. Flouzat,[a] O. Gevin,[a] F. Guilloux,[a] H. Hernandez,[f] I. Mandjavidze,[a] M. Munhoz,[b] B. Sanches,[c] L. Severo,[e] J.N. Soares,[d] W. Van Noije,[c] and F. William da Costa[c]

[a] *CEA IRFU, Université Paris-Saclay,*
*route de Saclay, 91191 Gif sur Yvette Cedex, France*
*E-mail:* damien.neyret@cea.fr

[b] *Instituto de Física da Universidade de São Paulo*
*Rua do matão, 1371, 05508-090 São Paulo, Brazil*

[c] *Escola Politécnica da Universidade de São Paulo*
*Avenida Prof. Luciano Gualberto, Travessa do Politécnico, 380, 05508-010 São Paulo, Brazil*

[d] *Escola de Engenharia de São Carlos, Universidade de São Paulo*
*Avenida Trabalhador são-carlense, 400, 13566-590, São Carlos, Brazil*

[e] *Divisão Engenharia Eletrônica, Instituto Tecnológico de Aeronáutica,*
*Praça Marechal Eduardo Gomes, 50 Vila das Acácias, 12228-900 São José dos Campos, Brazil*

[f] *TID Instrumentation Division, Integrated Circuits Department, SLAC*
*2575 Sand Hill Road, Menlo Park, CA 94025 Stanford University, CA, USA.*

ABSTRACT: The SALSA chip is a future readout ASIC foreseen for the MPGD detectors, developed in the framework of the EIC collider project, to equip the MPGD trackers of the EPIC experiment. It is designed to be versatile, to be adapted to other usages of MPGD detectors like TPC or photon detectors. It integrates a frontend block and an ADC for each of the 64 channels, associated to a configurable DSP processor meant to correct data and reduce the raw data flux to limit the output bandwidth. It will be compatible with the continuous readout foreseen for the EPIC DAQ, but will also work in a triggered environment. Several prototypes are already produced in order to qualify the different blocks of the chip, in particular the frontend, the ADC and the clock generation. The next 32-channel prototype is currently under development and is planed to be produced in 2025. The final prototype will be produced and tested from 2026 for a production of the SALSA chip at the horizon of 2027.

---

1 Corresponding author.

# 1. Introduction and context

The Electron-Ion Collider (EIC) [1] will be installed at the Brookhaven National Laboratory (BNL) in USA in early 2030s, with a 5-18 GeV electron beam colliding with a 40-275 GeV proton beam, which could also be ~100 GeV/nucleon ion beam. The EIC EPIC experiment [2] consists of a 2 T solenoid equipped with several layers of trackers and calorimeters to reconstruct the scattering angles and the momentum of particles, and RICH (Ring Imaging Cherenkov) and time-of-flight detectors for particle identification. The central tracker [3] is based on several layers of MAPS silicon detectors (Monolithic Active Pixel Sensors) surrounded by Micro-Pattern Gaseous Detectors (MPGD) (Figure 1). 2-D curved Micromegas detectors [4] form the external layer of the barrel while flat wheels of µ-RWell planes (micro-resistive well detector) cover both endcaps. µ-RWell planes are placed around the AC-LGAD time-of-flight detectors (capacitively-coupled Low-Gain Avalanche Diode) surrounding the barrel. All these MPGD detectors will be read by a new readout ASICs, the SALSA chip.

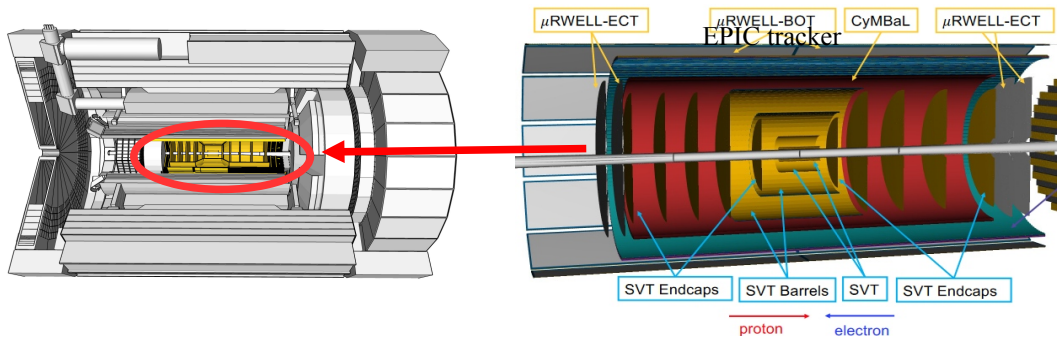

Figure 1: View of the central part of the EPIC experiment with the central tracker highlighted (left). Detail of the tracker, with the cylindrical Micromegas (CymBaL) in red and the µ-RWell in grey (right).

The SALSA readout electronics is designed to detect MPGD detector [5] signals, which are in the order of 50-100 ns long, with an charge of a few 10 fC up to ~250 fC. The setup will be exposed to a mild radiation environment, with doses in the order of 10 krad and $10^{11}$ n/cm² over a period of 10 years, and with a magnetic field of 2 T. The expected signal rate will be in the order of 10 kHz, with an expected detector time resolution in the order of 10 ns.

# 2. SALSA specifications and architecture

## 2.1 Specifications

The SALSA ASIC (Application-Specific Integrated Circuit) is expected to convert, amplify and shape analog MPGD signals from 64 channels, then to digitize these signals with integrated Analog to Digital Converters (ADC), and finally to process samples with an integrated digital signal processor (DSP). It is meant to be a versatile ASIC for a wide field of MPGD applications (trackers, TPC, photon detectors, etc...) and possibly other kinds of detectors. The peaking time, the gain, and the charge preamplifier can be configured to be adapted to the particular detector type. The preamplifier can be adjusted to up to 100 kHz counting rates, and an anti-saturation system is foreseen to prevent long dead times and saturation-induced cross-talk. The ADC sampling rate can be adapted to the width of the shaped analog signals. The TSMC 65 nm technology is chosen for its improved performances and sustainability compared to older 130 nm technologies.

The project is a common initiative of the University of Sao Paulo (USP, Brazil) and the CEA Saclay IRFU (France), which developed several readout ASICs for MPGD and other detectors, namely the SAMPA chip for ALICE TPC [6] at USP, and the AFTER[7], AGET [8] and DREAM [9] MPGD frontend chips at IRFU. These institutes have complementary competences, on low-noise generic frontends at IRFU, and on integrated ADC and digital processing in chip at USP.

The specifications of the SALSA ASIC are listed in Table 1. They are voluntarily extended compared to the specific requirements for the EPIC MPGD detector electronics. The ASIC is compatible with the continuous readout DAQ architecture foreseen for EPIC, as well as a triggered acquisition environment.

| Characteristics | Value | Comments |
|---|---|---|
| Number of channels | 64 | |
| Input capacitance | 50-200 pF | Reasonable gain up to 1 nF |
| Charge ranges | 0-50 fC to 0-5 pC | 4 gain ranges |
| Peaking times | 50 to 500 ns | 8 values |
| Input rates | Up to 100 kHz/ch | Fast CSA reset |
| ADC dynamics | 12 bits | > 10 effective bits |
| Sampling rate | programmable in range 5-50 MS/s | |
| Reference clock | 40-100 MHz | |
| Data output | 4 x 1 Gb/s links | 1 to 4 can be activated |
| Die size | ~ 1 cm² | TSMC 65 nm technology |
| Power consumption | ~ 15 mW/channel | 1.2 V power voltage |

**Table 1.** Specifications of the SALSA ASIC

### 2.2 Architecture

The SALSA architecture (Figure 2) combines channel frontends and ADC, and a DSP to process the ADC samples. The frontend is based on a charge sensitive amplifier (CSA), a pole-zero cancellation circuit (PZC) and a shaper, able to read signals of both polarities. The CSA input stage is based on two transistors of different sizes (2 and 6 mm width, being folded in order to fit in the small die surface) put in parallel, with the possibility to disconnect the largest one to adapt that stage to detector capacitance and signal amplitudes. When the input capacitance is low, only the 2 mm transistor is used in order to limit the electronics noise in this configuration, while the 6 mm one is added for input capacitances larger than a few tens of pF as a larger transistor is better fitted for larger capacitances. The precise input capacitance value above which 6+2 mm transistors are better optimized than 2-mm only is yet to be determined.

Four charge ranges (gains) are implemented in the CSA: 0-50 fC, 0-250 fC, 0-500 fC or 0-5 pC. The shaping can be done with eight different peaking times from 50 to 500 ns (every 50 ns up to 350 ns, plus 500 ns). An anti-saturation circuit is integrated to recover frontend from large amplitudes in specific cases, like a spark occurring in the gaseous detector. It is based on a transistor discharging the CSA feedback capacitor, which is activated when the CSA output goes above a given threshold. The ADC is based on a successive approximation register (SAR ADC) on 12 bits at a maximum sampling rate of 50 MS/s.

The DSP processes data to correct the signal baseline and suppress the empty samples, applying pedestal equalization, a common mode noise correction, a baseline following correction, an infinite impulse filtering, and finally a low-amplitude sample suppression. A feature extraction module is foreseen to identify in the sample stream the peaks corresponding to effective hits in the detector channels, and extract their amplitude, time of arrival and signal width. External triggers are managed by the DSP to select specifically samples present in the trigger time window, and trigger primitives will be generated when samples above threshold will be observed, with conditions on channel multiplicities. A phased-lock loop (PLL) circuit generates internal clocks needed for the different functions of the chip, keeping them synchronized with the reference clock.

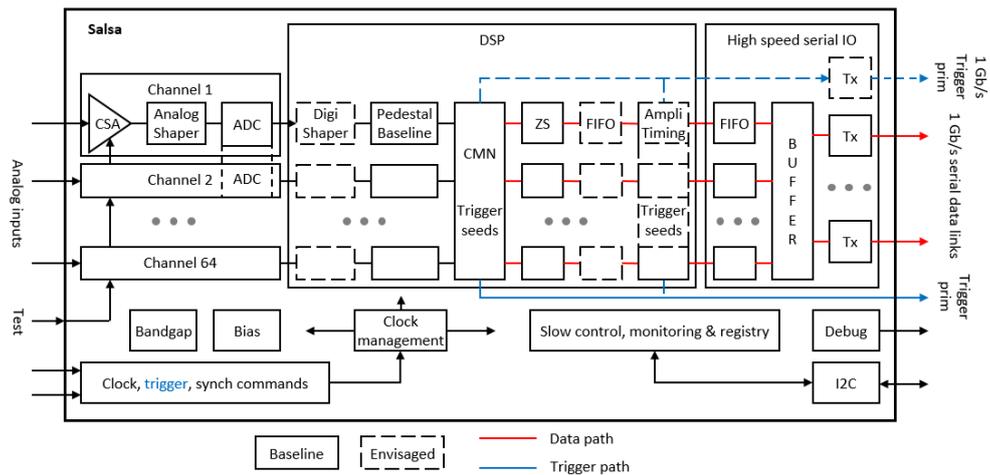

Figure 2: Architecture of the SALSA ASIC

## 3. Timeline and recent results

### 3.1 Timeline of the project

First SALSA0 prototypes were built beginning of 2023 (Figure 3 left) to test first versions of the frontend block and the ADC. The prototypes were produced using the TSMC 65nm "Mini@sic" program of the IMEC company associated with CERN. The PRISME prototype (Figure 3 center) was produced in 2023 in order to test specifically the new PLL block. The SALSA1 prototypes (Figure 3 right) were produced in 2024 to test improved frontend and ADC designs connected together. Several service blocks are also tested in the PRISME and SALSA1 chips. A 32-channel SALSA2 prototype will be submitted in 2025, with most of the DSP functionalities and high speed serial links in place. At last a final SALSA3 prototype is foreseen in 2026 with the nominal number of channels and a fully featured DSP. It will be a pre-series version of the ASIC before the final production.

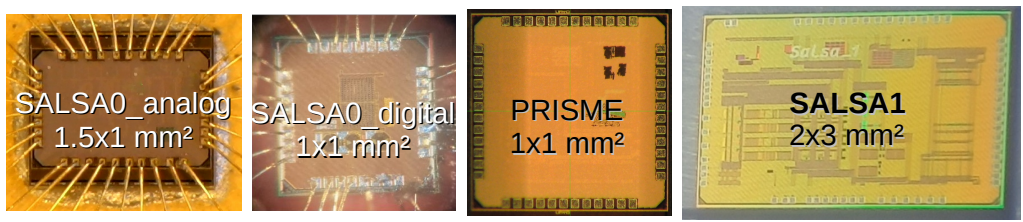

Figure 3: Pictures of the existing SALSA prototypes

## 3.2 Recent results

Different configurations were tested with the SALSA0 prototypes (Figure 4 left) to measure their performance and compare them with simulations. Equivalent noise charge dependency on input capacitance was as expected, showing only 1200 e⁻ for typical MPGD application (Figure 4 right). Fast recovery tests with anti-saturation circuit also showed that a 50 pC input charge was evacuated in around 50 μs (Figure 5). Tests with PRISME showed that the nominal 3.2 GHz PLL frequency was reached with a wide input frequency range, and clock signals were generated with a random jitter of 2.5 ps RMS. An updated prototype design PRISMEv1 correcting some unexpected deterministic jitter was submitted in December 2024.

## 4. Conclusions and prospects

The SALSA ASIC is a versatile readout chip currently under development for MPGD detectors, offering large ranges of gain, peaking times, input capacitance and sampling rates, able to work with the triggerless readout DAQ foreseen at EPIC. The ASIC will group a frontend and an ADC block for each of the 64 channels, and a DSP to process and reduce these data. The SALSA2 prototype development is presently ongoing, with an important work in progress on the design of the DSP, with specifications mostly finalized. Tests on SALSA1 prototype are starting. The timeline of the project is compatible with the schedule of the EPIC experiment, with a production of the readout ASICs at the horizon of 2027. Other projects also expressed interest in this chip, thanks to its versatility aspects.

## Acknowledgments

This work is supported by the Generic EIC-Related Detector R&D Program, and by the eRD109 EIC Project R&D Program. The development of the 32-channels prototype and beyond is also supported by the Agence Nationale de la Recherche (ANR) and the Fundação de Amparo à Pesquisa do Estado de São Paulo (FAPESP) under the contract numbers ANR-24-CE31-7003-01 and 2024/04802-9 – ANR. The authors thank the CERN micro-electronics group for the authorization to reuse some of the 65nm IP blocks developed by this group.

Transactions on Nuclear Science 55.3 (2008): 1744-1752

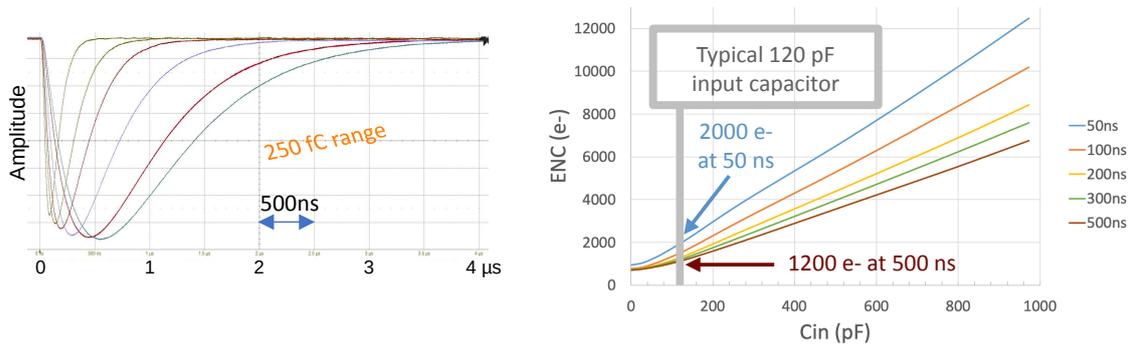

Figure 4: Response of the frontend for different peaking times from 50 to 500 ns with the 250 fC gain range (left). Equivalent noise charge vs input capacitance at different peaking times (right).

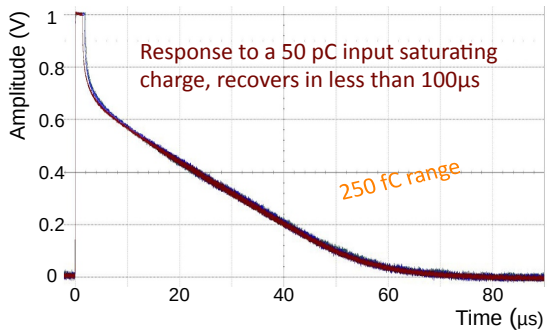

Figure 5: Response of the CSA to a large input charge, illustrating the role of the anti-saturation circuit